\documentclass[11pt]{article} 
\usepackage{moriond,epsfig}

\def\PRL{\em Phys. Rev. Lett.} 
 
\def \AAP #1 #2 {{\em Astron. Astrophys.\/} {\bf #1}, #2}   
\def \AAL #1 #2 {{\em Astron. Astrophys. Lett.\/} {\bf #1}, L#2}   
\def \AAR #1 #2 {{\em Astron. Astrophys. Rev.\/} {\bf #1}, #2}   
\def \AAS #1 #2 {{\em Astron. Astrophys. Suppl. Ser.\/} {\bf #1}, #2}   
\def \AJ #1 #2 {{\em Astron. J.\/} {\bf #1}, #2}   
\def \ANNREV #1 #2 {{\em Ann. Rev. Astron. Astrophys.\/} {\bf #1}, #2}   
\def \APJ #1 #2 {{\em Astrophys. J.\/} {\bf #1}, #2}   
\def \APJL #1 #2 {{\em Astrophys. J. Lett.\/} {\bf #1}, L#2}   
\def \APJS #1 #2 {{\em Astrophys. J. Suppl.\/} {\bf #1}, #2}   
\def \APSS #1 #2 {{\em Astrophys. Space Sci.\/} {\bf #1}, #2}   
\def \ASR #1 #2 {{\em Adv. Space Res.\/} {\bf #1}, #2}   
\def \MN #1 #2 {{\em Mon. Not. R. Astr. Soc.\/} {\bf #1}, #2}   
\def \PRL #1 #2 {{\em Phys. Rev. Lett.\/} {\bf #1}, #2}   
\def \NAT #1 #2 {{\em Nature\/} {\bf #1}, #2}

\def\be{\begin{equation}} 
\def\ee{\end{equation}} 
\def\bea{\begin{eqnarray}} 
\def\eea{\end{eqnarray}} 
 
\begin{document} 
 
\vspace*{4cm} 
 
\title{NUMERICAL SIMULATIONS OF RELATIVISTIC SHOCK ACCELERATION} 
   
\author{Micha{\l} Ostrowski} 
   
\address{Obserwatorium Astronomiczne, Uniwersytet Jagiello\'nski, 
ul. Orla 171, 30-244 Krak\'ow, Poland (E-mail:  mio{@}oa.uj.edu.pl)}   
   
\maketitle 
\abstracts{ 
We review the present status of the cosmic ray acceleration theory in 
mildly relativistic shock waves. Due to the involved substantial 
particle anisotropies analytical methods can tackle only simple 
situations involving weakly turbulent conditions near the shock. The 
numerical Monte Carlo methods are used to study the acceleration process 
in more general conditions. Contrary to non-relativistic 
shocks, the cosmic ray spectra at relativistic shock waves depend 
substantially on the magnetic field configurations and turbulence 
spectra. Depending on these conditions both very flat and very steep 
spectra can be created, with the characteristic acceleration time scales 
changing non-monotonously with the turbulence amplitude. Thus, the 
discussed theory is not able to uniquely explain the observed 
synchrotron spectra of relativistic shocks. We also mention an 
interesting possibility of particle acceleration at incompressible flow 
discontinuities (shear layers) at boundaries of relativistic jets. }

\section{Introduction}   
   
Relativistic plasma flows are detected or postulated to exist in a 
number of astrophysical objects, ranging from a mildly relativistic jet 
of SS433, through the-Lorentz-factor-of-a-few jets in AGNs and galactic 
`micro-quasars', up to ultra-relativistic outflows in sources of gamma 
ray bursts and, possibly, in pulsar winds. As nearly all such objects 
are efficient emitters of synchrotron radiation and/or high energy 
photons, our attempts to understand the processes generating cosmic ray 
particles are essential for understanding the fascinating phenomena 
observed. Below we will discuss the work carried out over the last 15 
years in order to understand the cosmic ray acceleration processes 
acting at relativistic shocks. We limit the discussion to mildly 
relativistic flows with Lorentz factors up to, say, 10. The acceleration 
processes acting at non-relativistic and ultra-relativistic shocks are 
discussed in this proceedings by Gallant. We also shortly discuss a 
possibility for cosmic ray acceleration at shear layers accompanying 
relativistic jets.

\section{Cosmic ray acceleration at relativistic shock waves}   
   
\subsection{The Fokker-Planck description of the acceleration process}   
   
In the case of the shock velocity reaching values comparable to the 
light velocity, the particle distribution at the shock becomes 
anisotropic. It complicates to a great extent both the physical picture 
and the mathematical description of particle acceleration. The first 
attempt to consider the acceleration process at the relativistic shock 
was presented in 1981 by Peacock, however, no consistent theory was 
proposed until a paper of Kirk \& Schneider (1987a). Those 
authors considered the stationary solutions of the relativistic 
Fokker-Planck equation for the case of a parallel shock wave. For the 
applied momentum pitch-angle diffusion operator, $\partial / \partial 
\mu (D_{\mu \mu} \partial f / \partial \mu)$, they generalised the 
diffusive approach to higher order terms in particle distribution 
anisotropy and constructed general solutions at both sides of the shock 
which involved solutions of the eigenvalue problem. By matching two 
solutions at the shock, the spectral index of the resulting power-law 
particle distribution can be found by taking into account a sufficiently 
large number of eigenfunctions. The same procedure yields the particle 
angular distribution and the spatial density distribution. The low-order 
truncation in this approach corresponds to the standard diffusion 
approximation and to a somewhat more general method described by 
Peacock. The above analytic approach (or the `semi-analytic' one, as the 
mentioned matching of two series involves numerical fitting of the 
respective coefficients) was verified by Kirk \& Schneider (1987b) by 
the method of particle Monte Carlo simulations. 
   
An application of this approach to more realistic conditions -- but 
still for parallel shocks -- was presented by Heavens \& Drury (1988), 
who investigated the fluid dynamics of relativistic shocks (cf. also 
Ellison \& Reynolds 1991) and used the results to calculate spectral 
indices for accelerated particles. They considered the shock wave 
propagating into electron-proton or electron-positron plasma, and 
performed calculations using the analytic method of Kirk \& Schneider 
for two different power spectra for the scattering MHD waves. In 
contrast to the non-relativistic case, they found that the particle 
spectral index\footnote{We use here a phase space distribution function 
index $\alpha$ and the synchrotron spectral index $\gamma = (\alpha - 
3)/2$. An index `1' (`2') is used for quantities measured in the plasma 
rest frame upstream (downstream) of the shock. In the text, $U$ is a shock 
velocity, $\Psi$ is an inclination of the magnetic field, ${\bf B}$, 
with respect to the shock normal, $U_{B,1} \equiv U_1/\cos{\Psi_1}$ is 
the shock velocity projected at the magnetic field. An amplitude of the 
magnetic turbulence, $\delta B$, is often characterized with the ratio 
of the particle cross-field diffusion coefficient, $\kappa_\perp$, to 
the parallel diffusion coefficient, $\kappa_\parallel$. }, $\alpha$, 
depends on the form of the wave spectrum.  The unexpected fact was noted 
that the non-relativistic expression $\alpha = 3R/(R-1)$ ($R$ - a shock 
compression ratio) provided a quite reasonable approximation to the 
actual spectral index (see the initial points of curves at Fig.~1, the 
ones for $\Psi_1 = 0$, or $U_1 = U_{B,1}$). 
   
\begin{figure}[hbt]   
\vspace{82mm} 
\includegraphics{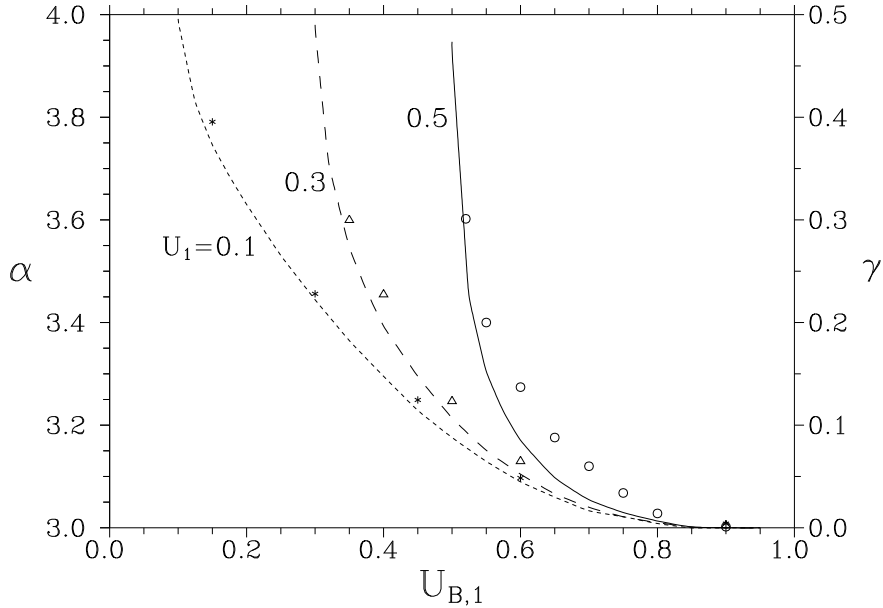} 
\vspace{0mm} 
\caption{  
Spectral indices $\alpha$ of particles accelerated at oblique shocks   
versus shock velocity projected at the mean magnetic field, $U_{B,1}$.   
On the right the respective synchrotron spectral index $\gamma$ is   
given. The shock velocities $U_1$ are given near the respective curves   
taken from Kirk \& Heavens (1989). The points were taken from   
simulations deriving explicitly the details of particle-shock   
interactions (Ostrowski 1991a). The results are presented for   
compression $R = 4$.} \end{figure}   
   
A substantial progress in understanding the acceleration process with 
a highly anisotropic particle distribution at the shock 
brought the work of Kirk \& Heavens (1989; see also Ostrowski 1991a 
and Ballard \& Heavens 1991), who considered particle acceleration at 
{\it subluminal} ($U_{B,1} < c$) relativistic shocks with oblique 
magnetic fields. They assumed the magnetic momentum conservation, 
$p_\perp^2/B = const$, at particle interaction with the shock and 
applied the Fokker-Planck equation discussed above to describe particle 
transport along the field lines outside the shock, while excluding  the 
possibility of cross-field diffusion. In the cases when $U_{B,1}$ 
reached relativistic values, they derived very flat energy spectra with 
$\gamma \approx 0$ at $U_{B,1} \approx c$ (Fig.~1). In such conditions, 
the particle density in front of the shock can substantially -- even by 
a few orders of magnitude -- exceed the downstream density (see the 
curve denoted `-8.9' at Fig.~2). Creating flat spectra and great density 
contrasts is due to the effective reflections of anisotropically 
distributed upstream particles from the region of compressed magnetic 
field downstream of the shock. 
 
As stressed by Begelman \& Kirk (1990), in relativistic shocks one can   
often find the {\it superluminal} conditions with $U_{B,1} > c$, where   
the above presented approach is no longer valid. Then, it is not   
possible to reflect upstream particles from the shock and to transmit   
downstream particles into the upstream region. In effect, only a single   
transmission of upstream particles re-shapes the original distribution   
by shifting particle energies to larger values. The energy gains in such   
a process, involving a highly anisotropic particle distribution at the 
shock, can be quite significant, exceeding the value expected for the 
adiabatic compression. 
   
The approach proposed by Kirk \& Schneider (1987a) and Kirk \& Heavens 
(1989), and the derivations of Begelman \& Kirk (1990) are valid only in 
case of weakly perturbed magnetic fields. However, if finite amplitude 
MHD waves are present in the medium, both the Fokker-Planck approach is 
no longer valid and the approximate magnetic momentum conservation no 
longer holds for particles interacting with oblique shocks. Then 
numerical methods have to be used.

\subsection{Particle acceleration in the presence of finite amplitude 
magnetic field perturbations}   
   
\begin{figure}[hbt]   
\vspace{86mm} 
\includegraphics{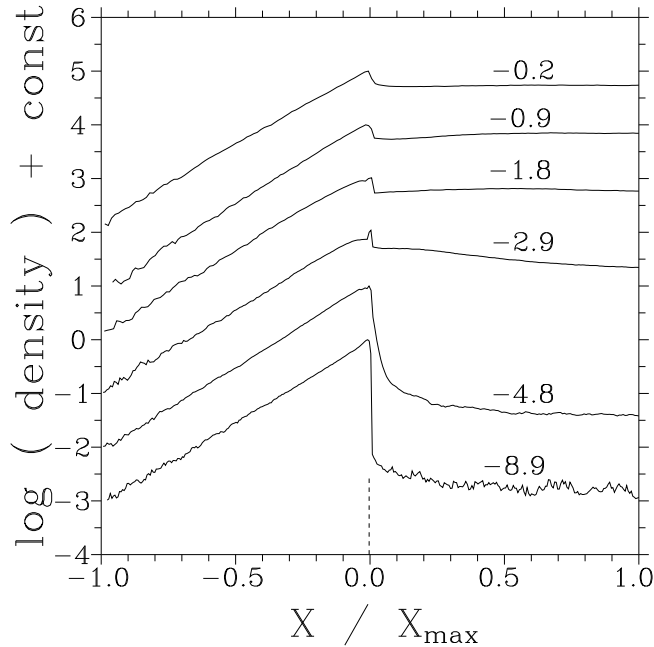} 
\vspace{0mm} 
\caption{  
The energetic particle density across the relativistic shock   
with an oblique magnetic field (Ostrowski 1991b). The shock with $U_1 =   
0.5 c$, $R = 5.11$ and $\psi_1 = 55^\circ$ is considered. The curves for 
different perturbation amplitudes are characterized with the value of 
$\log \kappa_\perp / \kappa_\parallel$ given near the curve. The data 
are vertically shifted for picture clarity. The value $X_{max}$ is the 
distance from the shock at which the upstream particle density decreases 
to $10^{-3}$ part of the shock value.} 
\end{figure} 
   
The first attempt to consider the acceleration process at a parallel shock 
wave propagating in a turbulent medium was presented by Kirk \& 
Schneider (1988), who included into the kinetic transport equation the 
Boltzmann collision operator describing the large angle scattering. By 
solving the resulting integro-differential equation they demonstrated 
the hardening of the particle spectrum for increasing contribution of 
the large-angle scattering. The reason for such a spectral change is the 
additional isotropization of particles interacting with the shock, 
leading to an increase in the particle mean energy gain. In oblique 
shocks, this simplified approach cannot be used because the character of 
individual particle-shock interaction -- reflection and transmission 
characteristics -- depends on the magnetic field perturbations. Let us 
additionally note that application of the {\em point-like large-angle} 
scattering model in relativistic shocks does not provide a viable 
physical representation of the scattering at MHD waves (Bednarz \& 
Ostrowski 1996). 
 
To handle the problem of the particle spectrum in a wide range of 
background conditions, the Monte Carlo particle simulations were widely 
used (e.g. Kirk \& Schneider 1987b; Ellison et al. 1990; Ostrowski 
1991a, 1993; Ballard \& Heavens 1992, Naito \& Takahara 1995, Bednarz \& 
Ostrowski 1996). At first, let us consider subluminal shocks. The field 
perturbations influence the acceleration process in a few ways. As 
they enable the particle cross field diffusion, a modification 
(decrease) of the downstream particle escape probability allows larger 
number of particles to re-cross the shock upstream. This factor tends to 
harden the spectrum. Next, the perturbations decrease particle 
anisotropy, leading to an increase of the mean energy gain of reflected 
upstream particles, but -- what is more important for oblique shocks -- 
they also increase the particle upstream-downstream transmission 
probability due to less efficient reflections, enabling additional 
particles to escape from further acceleration. The third factor is due 
to perturbing particle trajectory during an individual interaction with 
the shock discontinuity and breakdown of the approximate conservation of 
$p_\perp^2/B$. Because reflecting a particle from the shock requires a 
fine tuning of the particle trajectory with respect to the shock 
surface, even small amplitude perturbations can decrease the reflection 
probability in a substantial way (cf. Ostrowski 1991). Simulations show 
(see Fig.~3 for $U_{B,1} \le 1.0$, $\Psi_1 \le 60^\circ$) that -- until 
the wave amplitude becomes very large -- the factors leading to 
efficient particle escape dominate with the resulting steepening of the 
spectrum to $\gamma \sim 0.2$ -- $0.8$ ($\alpha \ge 3.4$), and the 
increased downstream transmission probability lowers the cosmic ray 
density contrast across the shock (Fig.~2). 
   
In parallel shock waves propagating in a highly turbulent medium, the   
effects discovered for oblique shocks can also manifest their  presence 
because of the {\it local} perturbed magnetic field compression at the 
shock. The problem was considered using the technique of particle 
simulations by Ballard \& Heavens (1992; cf. Ostrowski 1988 for a 
non-relativistic shock). They obtained very steep spectra in the 
considered cases, with the synchrotron spectral index growing from 
$\gamma \sim 0.6$ at medium relativistic velocities up to nearly $2.0$ 
at $U_1 = 0.98 c$. These results apparently do not correspond to the 
large-perturbation-amplitude limit of Ostrowski's (1993; see the 
discussion therein) simulations for oblique shocks and the analytic 
results of Heavens \& Drury (1988). 
   
For large amplitude magnetic field perturbations the acceleration   
process in superluminal shocks can lead to the power-law particle 
spectrum formation (cf. Fig.~3 for $\Psi_1 = 72.5^\circ$), against the 
statements of Begelman \& Kirk (1990) valid at small wave amplitudes 
only. Such  a general case was discussed by Ostrowski (1993) and by 
Bednarz \& Ostrowski (1996, 1998).

\subsection{The acceleration time scale}   
   
The shock waves propagating with relativistic velocities raise a 
question about the involved cosmic ray acceleration time scales, 
$T_{acc}$. A simple comparison to the values derived with the 
non-relativistic formula shows that $T_{acc}$ relatively decreases with 
increasing shock velocity for parallel (Quenby \& Lieu 1989; Ellison et 
al. 1990) and oblique (Takahara \& Terasawa 1990; Newman et al. 1992; 
Lieu et al. 1994; Quenby \& Drolias 1995; Naito \& Takahara 1995) 
shocks. However, the numerical approaches used in the listed papers, 
based on assuming particle isotropization for all scatterings, neglect 
or underestimate a significant factor affecting the acceleration process 
-- the particle anisotropy. Ellison et al. (1990) and Naito \& Takahara 
(1995) included also the more realistic derivations involving the 
pitch-angle diffusion approach. The calculations of Ellison et al. for 
parallel shocks show similar results to those they obtained for large 
amplitude scattering. For the shock with velocity $0.98 c$ the 
acceleration time scale is reduced by the factor $\sim 3$ with respect 
to the non-relativistic formula. Naito \& Takahara considered shocks 
with oblique magnetic fields. They confirmed the reduction of the 
acceleration time scale with increasing inclination of the magnetic 
field, the fact derived earlier for non-relativistic shocks. However, their 
approach neglected effects of particle cross field diffusion and assumed 
the adiabatic invariant conservation at particle interactions with the 
shock, thus limiting the validity of their results to a small amplitude 
turbulence near the shock. 
   
\begin{figure}[hbt]   
\vspace{82mm} 
\includegraphics{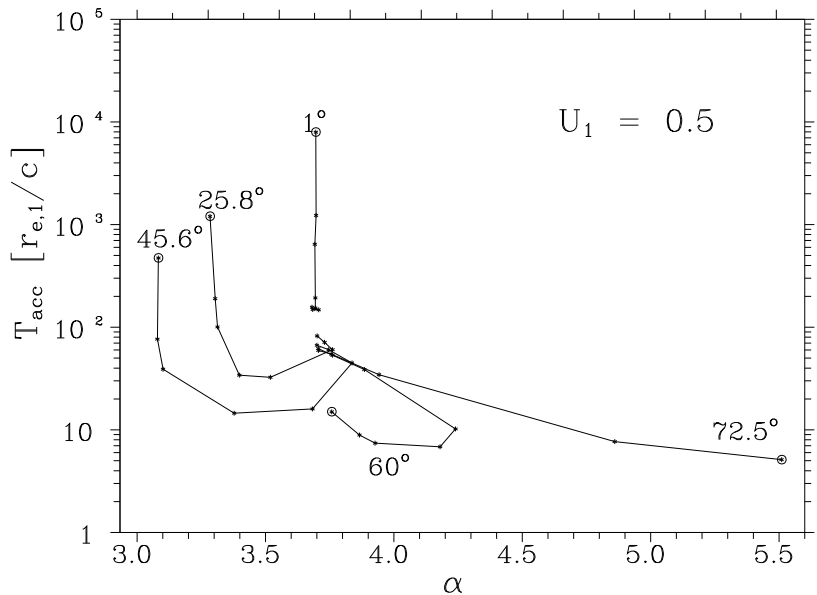} 
\vspace{0mm} 
\caption{  
The relation of $T_{acc}$ versus the particle spectral index   
$\alpha$ at different magnetic field inclinations $\psi_1$ given near   
the respective curves (from Bednarz \& Ostrowski 1996). A turbulence 
amplitude grows along the presented curves. The {\it 
minimum} value of the model parameter $\kappa_\perp/\kappa_\|$ occurs at 
the encircled point of each curve and the wave amplitude monotonously 
increases along each curve up to $\delta B \sim B$. $r_{e,1}$ is the 
particle gyroradius in the effective (including perturbations) upstream 
magnetic field.} 
\end{figure} 
   
A wider discussion of the acceleration time scale is presented by   
Bednarz \& Ostrowski (1996), who apply numerical simulations involving   
the small angle particle momentum scattering. The approach treats 
properly the existing correlations between the particle energy gains and 
the respective particle diffusion times off the shock. It is also 
believed to provide a reasonable description of particle transport in 
the presence of large turbulence amplitude and 
enables modelling of the cross-field diffusion effects. The resulting 
acceleration time scales given in the shock {\it normal} rest frame (cf. 
Begelman \& Kirk 1990) are presented on Fig.~3 versus the respective 
spectral index $\alpha$. The results for varying turbulence amplitudes 
are given for several magnetic field inclinations. In parallel ($\Psi_1 
= 1^\circ$) shocks $T_{acc}$ diminishes with the growing perturbation 
amplitude and -- not presented on the figure -- the growing shock 
velocity. However, it is approximately constant for a given value of 
$U_1$, if we use the formal diffusive time scale, $\kappa_{\parallel , 
1}/(U_1c) + \kappa_{\parallel , 2}/(U_2c)$, as the time unit. A new 
feature discovered in oblique shocks is a non-monotonic variation of 
$T_{acc}$ with the turbulence amplitude $\delta B$. The acceleration 
process leading to the power-law spectrum is possible in superluminal 
shocks only in the presence of large amplitude turbulence. Then, in 
contrast to the quasi-parallel shocks, $T_{acc}$ increases with 
increasing $\delta B$, accompanied with a substantial flattening of the 
spectrum. In the considered cases with the oblique field configurations 
one may note a possibility to have the extremely short acceleration time 
scale, comparable to the particle gyroperiod in the magnetic field 
upstream of the shock. 
   
\section{Energetic particle acceleration at relativistic shear layers} 
        
The acceleration processes acting in shocks formed in relativistic jets 
are not always able to explain the observed high energy electrons 
radiating far away from the shock. A natural but mostly unexplored 
possibility to explain such observations is the one involving particle 
acceleration at the interface between the jet and the surrounding it 
`cocoon'. To date the knowledge of physical conditions within such 
velocity shear layers is very limited and only rough estimates for the 
considered acceleration processes are possible. Within the jet turbulent 
boundary layer with a small velocity shear the ordinary second-order 
Fermi acceleration, as well as the process of `viscous' particle 
acceleration (cf. the review by Berezhko 1990 of the work done in early 
80-th; also Earl et al. 1988, Ostrowski 2000) can take place. A 
mean particle energy gain in the viscous acceleration processes scales 
as $<\Delta E> \, \propto \left( { <\Delta U> \over c } \right)^2 $, 
where $< \Delta U >$ is the mean velocity difference between the 
`successive scattering acts'. It is proportional to the mean free path 
normal to the shear layer, $\lambda_n$, times the mean flow velocity 
gradient in this direction $ \nabla_n \cdot \vec{U} $. With $d$ denoting 
the shear layer thickness this gradient can be estimated as $| \nabla_n 
\cdot \vec{U} | \approx U/d$. Because the acceleration rate in the Fermi 
II process is $\propto (V / c)^2$ ($V$  is the wave velocity $\approx$ 
the Alfv\'en velocity), the relative importance of both processes is 
given by a factor $(\lambda_n U \ d V )^2$. The relative efficiency of 
the viscous acceleration grows with $\lambda_n$ and in the formal limit 
of $\lambda_n \approx d$ it will dominate over the Fermi acceleration to 
a large extent. Because accelerated particles can escape from the 
accelerating layer only due to a relatively inefficient radial 
diffusion, the resulting particle spectra should be very flat, depending 
however on several unknown physical parameters of the boundary layer 
(Ostrowski 1990, 1998). Spectra of {\it very high} energy particles, 
with $\lambda_n \ge d$ (or $r_g > d$), do not require so detailed 
information about the boundary layer structure and can thus be derived 
with higher credibility. 
 
As an example let us present our results (Ostrowski 1998) of such 
numerical Monte Carlo derivations, comparing the ultra high energy 
particle spectrum generated at the jet terminal shock with the spectrum 
resulting from the jet boundary acceleration. The distributions of 
particles accelerated at the jet side-boundary, far from the shock, can 
be very flat. This feature results from the character of the 
acceleration process with particles having a chance to meet the 
accelerating surface and gain energy again and again for inefficient 
diffusive particle escape to the sides. Contrary to that, the shock 
acceleration process determines the spectrum inclination due to the 
joint action of the particle energization at the shock and the 
continuous escape due to particle advection with the downstream plasma. 
In the discussed simulations the escape probability grows with the 
particle energy providing a cut-off in the spectrum. For the shock 
spectrum, in the range of particle energies directly preceding the 
cut-off, the spectrum exhibits some flattening with respect to the 
inclination expected for the pure shock acceleration. There are two 
reasons for that flattening: additional particle transport from the 
downstream region to the upstream one through the cocoon surrounding the 
jet and inclusion of the flat spectral component resulting from the side 
boundary acceleration.

\section{Final remarks}   
   
The work done to date on the {\it test particle} cosmic ray acceleration   
at mildly relativistic shocks do not provide results allowing for 
meaningful modelling of the observed astrophysical sources. The main   
reason for that deficiency is -- in contrast to non-relativistic shocks   
-- a direct dependence of the derived spectra on the conditions at the   
shock. Not only the shock compression ratio, but also other parameters,   
like the mean magnetic field inclination or the wave spectrum 
and amplitude, are significant here. Depending on the actual 
conditions one may obtain spectral indices as flat as $\alpha = 3.0$   
($\gamma = 0.0$) or quite steep ones with $\alpha > 5.0$ ($\gamma > 
1.0$). The background conditions leading to the very flat spectra are   
probably subject to some instabilities; however, there is no detailed   
derivation describing the instability growth and the resulting cosmic   
ray spectrum modification. The situation may become simpler for   
ultra-relativistic shocks, as discussed by Gallant in this proceedings. 
   
A true progress in modelling particle acceleration in actual sources   
requires a full plasma non-linear description, including feedback of   
accelerated particles at the turbulent wave fields near the shock wave,   
flow modification caused by the cosmic rays' plasma pre-shock   
compression and, of course, the appropriate boundary conditions. A   
simple non-linear approach to the parallel shock case was presented by 
Baring \& Kirk (1990), who found that relativistic shocks could be very 
efficient accelerators. However, it seems to us that in a more general 
case it will be very difficult to make any substantial progress in that 
matter. For very flat particle spectra the non-linear acceleration 
picture depends to a large extent on the detailed knowledge of the 
background and boundary conditions in the scales relevant for particles 
near the upper energy cut-off. The existence of stationary solutions is 
doubtful in this case. An important step toward considering detailed 
physics of the acceleration provides the work of Hoshino et al. (1993) 
for ultra-relativistic shocks, but there is a lack of analogous studies 
for mildly relativistic shocks. 
   
One may note that observations of possible sites of relativistic shock   
waves (knots and hot spots in extragalactic radio sources), which allow   
for determination of the energetic electron spectra, often yield 
particle spectral indices close to $\alpha = 4.0$ ($\gamma = 0.5$). In   
order to overcome difficulties in accounting for these data Ostrowski   
(1994) proposed an additional {\it `law of nature'} for non-linear   
cosmic ray accelerators. The particles within different energy ranges do   
not couple directly with each other and are supposed to form independent   
`degrees of freedom' in the system. Our `law' provides that nature   
prefers energy equipartition between such degrees of freedom, yielding   
the spectra with $\alpha \approx 4.0$~.   
   
The acceleration processes at astrophysical shear layers may provide a 
viable explanation for some `strange' observational data in relativistic 
jets. For example it may have significant consequences for the 
relativistic jet structure and the high energy radiation of AGNs 
(Ostrowski 2000; Stawarz \& Ostrowski, in preparation). Also, these 
mechanisms could be responsible for accelerating nuclei up to the ultra 
high energy scales (Ostrowski 1998). The discussed processes are 
particularly interesting because a simple inspection does not reveal any 
physical obstacles which could make them inefficient. Unfortunately, the 
resulting particle spectra depend to a large extent on the poorly known 
background physical conditions. 
      
\section*{Acknowledgements} 
The present work was supported by the {\it Komitet Bada\'n Naukowych}   
through the grant PB 258/P03/99/17.

\end{document}